\documentclass[prl,twocolumn,floatfix]{revtex4}
\usepackage{graphics}
\usepackage{graphicx}
\usepackage{amsfonts}
\usepackage{amssymb}

\def\rti{R$_2$Ti$_2$O$_7$}
\def\tbti{Tb$_2$Ti$_2$O$_7$}

\def\tb{Tb$^{3+}$}

\begin{document}
\title{Anisotropic propagating excitations and quadrupolar effects in Tb$_2$Ti$_2$O$_7$} 
  
\author{Sol\`ene Guitteny$^1$, Julien Robert$^1$, Pierre Bonville$^2$, Jacques Ollivier$^4$, 
Claudia Decorse$^3$, Paul Steffens$^4$, Martin Boehm$^4$, Hannu Mutka$^4$, 
Isabelle Mirebeau$^1$ and Sylvain Petit$^1$} 
 
\affiliation{$^1$ Laboratoire L\'eon Brillouin, CEA Saclay, bât.563 91191 Gif-sur-Yvette Cedex} 
\affiliation{$^2$ DSM/IRAMIS/SPEC, CEA Saclay, 91191 Gif-sur-Yvette Cedex}
\affiliation{$^3$ ICMMO, Universit\'e Paris 11, 91405, Orsay, France}
\affiliation{$^4$ Institut Laue Langevin, 6 rue Jules Horowitz, BP 156 F-38042 Grenoble, France}

\begin{abstract}
The dynamical magnetic correlations in \tbti\, have been investigated using polarized inelastic 
neutron scattering. Dispersive excitations are observed, emerging from pinch points in reciprocal 
space and characterized by an anisotropic spectral weight. Anomalies in the crystal field and phonon 
excitation spectrum at Brillouin zone centers are also reported. These findings suggest that Coulomb 
phases, although they present a disordered ground state with dipolar correlations, allow the 
propagation of collective excitations. They also point out a strong spin-lattice coupling, which likely 
drives effective interactions between the $4f$ quadrupolar moments. 
\end{abstract}

\maketitle

%%%%%%%%%%%%%%%%%%%%%%%%%%%%%%%%%%%%%%%%%%%%%%%
%=============================================================

In the last decade, spin-ice physics in the \rti\ rare-earth pyrochlore, the celebrated lattice of corner 
sharing tetrahedra considered as the archetype of geometrical frustration in three dimensions, have 
aroused a lot of attention \cite{lhuillier,revgingras}. At the heart of this interest is a local constraint 
stating that each tetrahedron of the pyrochlore lattice must have, in its ground-state, two spins 
pointing in and two spins 
pointing out, the so-called "two-in two-out" ice rule, leading to a macroscopic degeneracy and to an 
emergent gauge structure \cite{isakov,henley}. The classical spins are quenched into one of the 
degenerate ground states formed by these configurations, resulting in an analog of water ice 
\cite{bramwell,castelnovo}. One of the main characteristics of this "Coulomb phase" is the existence 
of power law dipolar spin correlations, resulting in distinctive sharp and anisotropic features, the 
so-called "pinch-points", in neutron diffraction patterns \cite{henley}.

Other nontrivial states of matter may be produced in the quantum variant of spin ices. In this case, 
appreciable fluctuations between degenerate configurations are restored, resulting in a spin liquid 
state \cite{balents,molavian,onoda}. Current theoretical descriptions introduce a minimal Hamiltonian 
for pseudospins half, spanning the crystal electric field (CEF) ground doublet states $\vert \pm \rangle$, 
together with an Ising exchange constant $J_{zz}$ responsible for the spin-ice behavior, as well as 
"quantum" transverse terms $J_{\pm}$, $J_{z\pm}$ and $J_{\pm\pm}$ \cite{balents,savary}. 
For such large transverse terms, conventional phases are stabilized. They are characterized by a 
classical dipolar ordering in the case of Kramers ions, and by a quadrupolar ordering of the $4f$ 
quadrupoles for non-Kramers ions \cite{onoda,savary,sungbin}, accompanied by a coupling to the 
lattice degrees of freedom.
The quantum spin ice behavior is expected for moderate couplings and the ground state is a Coulomb 
phase described by an intricate superposition of "two-in two-out" configurations \cite{balents,molavian}. 
It exhibits exotic excitations with especially a two spinon continuum, as well as an emergent photon 
associated with the gauge structure \cite{shannon}. 

A potential candidate for quantum spin liquid is \tbti, which is characterized by an Ising-like anisotropy 
of the non-Kramers \tb\ ions along the local $\langle$111$\rangle$ axes \cite{molavian,cao}. In 
spite of effective antiferromagnetic interactions leading to a Curie-Weiss temperature of -13 K \cite{gingras}, 
which should drive the system into long-range order \cite{hertog,kao}, prior works pointed out a 
disordered fluctuating ground state down to 20 mK \cite{gardner99,gardner01}. Various subsequent
studies have suggested complex spin dynamics, where different time and temperature scales coexist, 
as revealed by muons \cite{gardner04,chapuis,yaouanc}, magnetization \cite{elsa,sarah} and neutron 
scattering experiments \cite{yasui,mirb07,rule07,rule09,petit,taniguchi,gaulin,takatsu,yin,fritsch}. Recently, 
power law spin correlations have also been reported \cite{fennell1}, bearing some resemblance with 
the pinch point pattern observed in aforementioned dipolar spin ices \cite{fennell2} and suggesting that 
the ground state of this material might be a Coulomb phase. 

To go further, we report in this letter a detailed description of the excitations 
emanating from this particular ground state. Combined elastic and inelastic neutron scattering 
measurements with polarization analysis provide evidence for the existence of low energy 
propagating excitations. Anomalies of the phonon modes, as well as of the first CEF level, are also 
observed, which unveil a strong dynamical coupling with the lattice.

%===========================================================

Low energy neutron experiments ($\hbar \omega \lesssim 0.5$~meV) were carried out on the 4F2 and 
IN14 triple axis spectrometers installed at LLB-Orphee (Saclay, France) and at the Institute Laue 
Langevin (Grenoble, France), respectively. The final energy was fixed to 3~meV, yielding an energy 
resolution $\Delta_0 \simeq 0.07$~meV (FWHM). Time-of-flight data were also collected on IN5 (ILL), 
with its recent single crystal set-up, with an incident wavelength $\lambda = 4$~\AA. 

%===========================================================

The magnetic correlations of several pyrochlore magnets have been studied in details 
by means of neutron diffraction \cite{fennell1}. Indeed, this technique provides a direct measurement of 
the spin pair correlation function $M(\mathbf{Q})=\sum_{i,j} \langle \mathbf{S}_{\perp,i}\mathbf{S}_{\perp,j} 
\rangle e^{-i\mathbf{Q}\cdot\mathbf{r}_{ij}}$, where $\mathbf{S}_{\perp,i}$ denotes the spin 
component at site $i$ perpendicular to the wavevector $\mathbf{Q}$. 
In dipolar spin ices, in which the spins are confined along the local easy axes $\langle 1 1 1 \rangle$, 
it has been possible to measure the usual "two-in two-out" correlations \cite{fennell1}, thanks to a clever 
experimental set-up combining the $(h,h,l)$ scattering plane and the use of spin polarization analysis. 
Those correlations are observed in the so-called "$M_y$" channel \cite{fennell2}.
In \tbti, the weaker anisotropy allows the spins to move away from their easy axes, resulting in 
additional correlations between the transverse spin components perpendicular to $\langle 1 1 1 \rangle$. 
The so-called "$M_z$" channel allows to measure those correlations (restricted however to spin components 
along the $z$ vertical axis $\parallel [1\bar 1 0]$), and points out antiferromagnetic "two-up two-down" 
spin configurations \cite{fennell2}.
Both kinds of correlations present pinch points at the Brillouin zone centers, but show maxima at different 
places in $\mathbf{Q}$-space \cite{fennell1}. For instance, the vicinity of $\mathbf{Q}=(2,2,0)$ is dominated 
by "two-up two-down'' transverse correlations (strong $M_z$) while $\mathbf{Q}=(1,1,1)$, is dominated by 
"two-in two-out'' like correlations (strong $M_y$).
\begin{figure}[t]
\includegraphics[width=8cm]{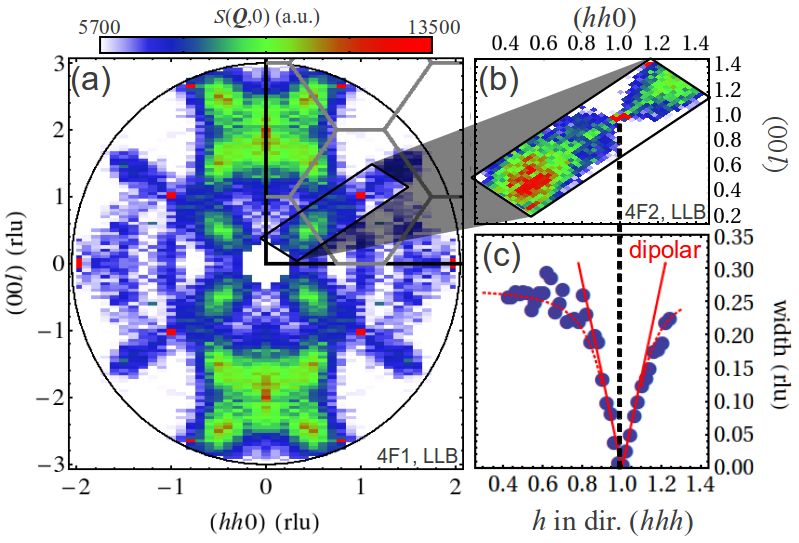}
\caption{(Color online): (a) map of the $\omega=0$ (elastic) scattering at $T=0.05$~K. The data within 
the top left corner have been actually measured and then symmetrized. Gray lines correspond to the 
boundaries of Brillouin zones. The data in figure (b) have been measured with smaller $\mathbf{Q}$-steps,  
focusing on the region close to the $(111)$ pinch point. (c) is a fit of the width along $q_\perp$ for different 
$q_\parallel$ varying along $\langle h,h,h \rangle$, as described in the text. }
\label{fig1}
\end{figure}
\begin{figure}[t]
\includegraphics[width=8cm]{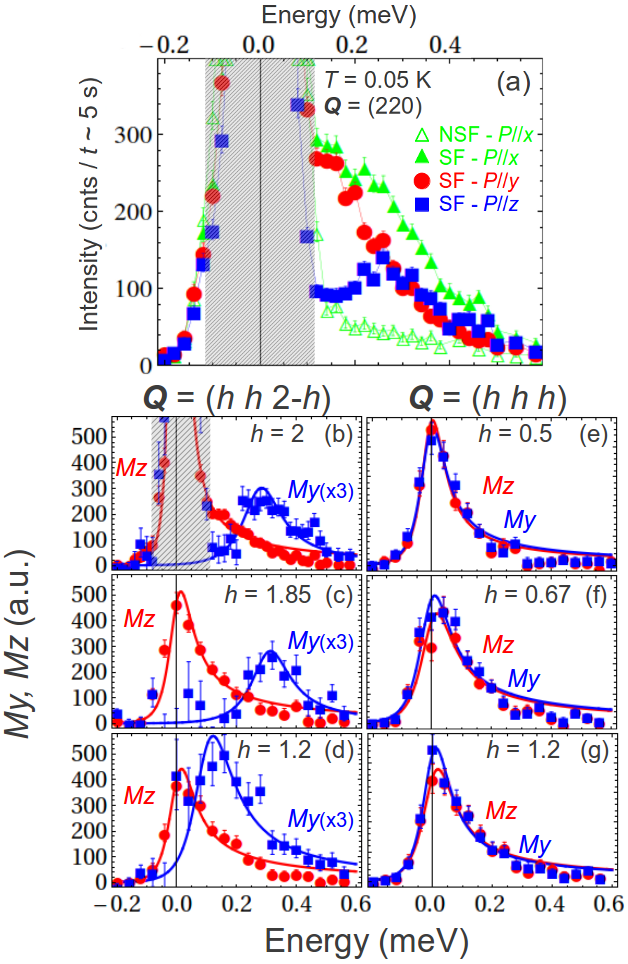}
\caption{(Color online) : (a) Spin Flip (polarization 
$\mathbf{P} \parallel \mathbf{x},\ \mathbf{y},\ \mathbf{z}$) and Non Spin Flip ($\mathbf{P}\parallel 
\mathbf{x}$) raw scans for $Q=(220)$ and $T=0.05$~K. 
The $\mathbf{x}$ $(\parallel \mathbf{Q})$, $\mathbf{y}$ $(\perp\mathbf{Q})$ in the scattering 
plane, and $\mathbf{z}$ $(\parallel[1\bar 10])$ axes are defined according to the spectrometer frame. 
(b,c,d) (resp. (e,f,g)) show $M_y$ and $M_z$ obtained by combining the different raw data \cite{regnault} 
with a flipping ratio $FR \simeq 20$ in direction $\langle h,h,2-h\rangle$ (resp. $\langle h,h,h\rangle$). 
The hatched areas hide the regions where the polarization analysis fails to suppress the nuclear 
background (Bragg peak contribution). The lines are the result of a fit as described in the text).}
\label{fig2}
\end{figure}
\begin{figure}[t]
\includegraphics[width=8cm]{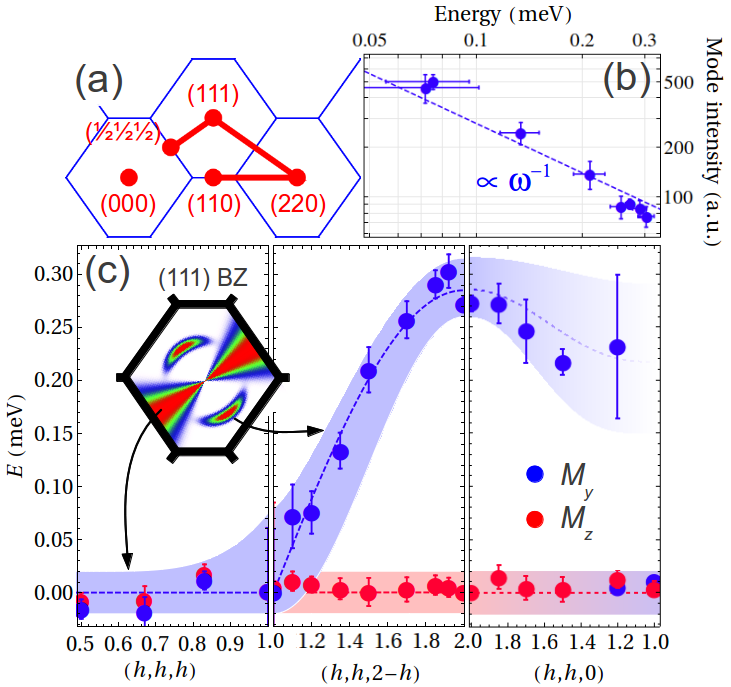}
\caption{(Color online) : (a) Sketch of the Brillouin zone indicating the directions of the scans carried 
out in the present study. (b) shows the evolution of the intensity of the dispersive inelastic mode as 
a function of $\omega$ along $\langle h,h,2-h\rangle$. (c) shows the dispersion of the mode along 
the three high symmetry directions. The inset is a schematic map of the $M_y$ spectral weight 
distribution in the Brillouin zone, which superimpose both the quasi-static (along $\langle h,h,h \rangle$) 
and the inelastic ("half moon") contributions.}
\label{fig3}
\end{figure}

Diffraction provides however an energy integrated response, so that energy resolved experiments, 
measuring 
$M(\mathbf{Q},\omega)=\int dt \sum_{i,j} \langle 
\mathbf{S}_{\perp,i} \mathbf{S}_{\perp,j}(t) \rangle 
e^{i(\omega t- \mathbf{Q}\cdot\mathbf{r}_{ij})}$ are important to further characterize 
the correlations. 
In this context, elastic $\omega=0$ data, obtained at $T=0.05$~K are shown in Figure \ref{fig1} (a). 
They are in qualitative agreement with the polarized data reported in Ref. \onlinecite{fennell2}. No magnetic 
Bragg peaks could be detected at $(1/2,1/2,1/2)$, as reported in Tb$_{2+x}$Ti$_{2-2x}$Nb$_x$O$_{7}$ 
\cite{ueland} and Tb$_{2+x}$Ti$_{2-x}$O$_{7+y}$ \cite{taniguchi}. This is consistent with the 
lattice parameter ($a=10.1528(5)$~\AA), determined precisely using x-ray scattering, and which 
positions our \tbti\, sample in the spin liquid phase \cite{taniguchi}. 
Focusing on the vicinity of $\mathbf{Q}=(2,2,0)$, a "two-up two-down" static correlation length of about 
$\xi=5\pm1$~\AA\, was determined by fitting the width of the corresponding pinch point in the $l$ direction 
(not shown). This value is comparable with the (energy integrated) diffraction results $\xi=2\pm0.2$~\AA.
A similar analysis around $\mathbf{Q}=(1,1,1)$ points out longer range "two-in two-out'' correlations, as 
seen on Figure \ref{fig1} (b). Following \cite{henley}, the structure factor close to the pinch point was fitted 
through a Lorentzian profile $q_\parallel^2/(q_\perp^2+q_\parallel^2+1/\xi^2)$, with $q_\parallel$ along 
$\langle h,h,h \rangle$, $q_\perp$ in the transverse direction and $\xi$ the static correlation length. An 
excellent agreement is obtained as shown in Fig. \ref{fig1} (c). The experimental width at the pinch point 
is limited by the instrument resolution, leading to $\xi>80$~\AA, at least one order of magnitude larger 
than the instantaneous one obtained from diffraction results ($\xi=8\pm 2$~\AA).These energy resolved 
data thus show that integrating over energy actually blurs very well defined pinch points, thus pointing 
out sizable magnetic fluctuations.

To further characterize the spectrum of these low energy fluctuations 
\cite{yasui,mirb07,rule07,rule09,petit,taniguchi,gaulin,takatsu,yin,fritsch}, polarized neutron experiments 
have been carried out on IN14, using the $M_{y,z}$ decomposition described above. Figure \ref{fig2} 
shows raw data taken at constant-$\mathbf{Q}=(220)$ (a), as well as $M_y$ and $M_z$ \cite{regnault} 
for different scattering vectors along high symmetry directions $\langle h,h,2-h \rangle$ (b,c,d) and 
$\langle h,h,h \rangle$ (e,f,g). These measurements show 
the existence of a dual response consisting of both an inelastic (blues squares) and a quasi-elastic (red disks) 
signal, as pointed out in \cite{petit}. The $M_z$ contribution is always found to be quasi-elastic. This is 
consistent with the very short range character of the "two-up two-down'' correlations, which may eliminate 
any possibility for coherent excitations to propagate.

In contrast, $M_y$ is different whether $\langle h,h,h \rangle$, $\langle h,h,2-h \rangle$ or 
$\langle h,h,0 \rangle$ is considered. Along $\langle h,h,h \rangle$, it is dominated by a 
quasi-elastic signal comparable to $M_z$ 
(Fig. \ref{fig2} (e-g)). Its intrinsic width (FWHM), roughly $\mathbf{Q}$-independent is around 
$\Gamma \simeq 0.15$~meV, providing a relaxation time $\tau \simeq$ 1.5~ps at $T=0.05$~K. 
Along $\langle h,h,2-h \rangle$ and $\langle h,h,0 \rangle$, $M_y$ shows gapless propagating 
excitations. The data has been fitted using a Lorentzian profile multiplied by the Bose factor and 
convoluted with the experimental resolution function. 
This provides the width, intensity and energy position of the mode reported in Figure \ref{fig3} (b) and (c). 
Stemming from the pinch point at $(1,1,1)$, it disperses significantly up to $\simeq 0.3$~meV at $(2,2,0)$, 
albeit more weakly along $\langle h,h,0 \rangle$. The presence of a small gap cannot be completely ruled 
out (at an energy however smaller than the experimental resolution $\Delta_0=0.07$~meV). Furthermore, 
as shown in Figure \ref{fig3} (b), the intensity of the mode along $\langle h,h,2-h \rangle$ decreases as 
$1/\omega$, a usual feature of magnetic excitations. This is very different from the behavior expected for 
the emergent photon, recently put forward in \cite{shannon}, and whose intensity is expected to grow as 
$\propto \omega$. A significant decrease of the spectral weight is also observed close to $(1,1,0)$. 
The propagation of such a collective excitation may be due to the spatial stiffness associated with the presence 
of algebraic correlations. The intrinsic width of the mode is however slightly larger than the resolution, a 
damping effect specific to systems having a strongly fluctuating ground-state \cite{chalker,Robert2008}.  

To illustrate $M_y$'s peculiar spectral weight distribution, the inset of Figure \ref{fig3} illustrates a 
constant energy cut taken in the vicinity of the pinch point at $(1,1,1)$. The feature 
along $(1,1,1)$ corresponds to quasi-static spin-ice-like correlations, while propagating 
excitations are visible along $\langle h,h,2-h \rangle$ and form the "half moon" features. 
This peculiar spectral weight distribution in reciprocal space can be understood by considering 
that the mode propagates defects which break the local constraint, hence giving rise to some response at 
positions in $\mathbf{Q}$-space which are in principle forbidden by the ice rule. Such observations have 
already been made numerically in the classical antiferromagnetic Heisenberg model on the pyrochlore 
\cite{Conlon2009,RobertUnpublished}, Kagome \cite{Robert2008}, and checkerboard \cite{RobertUnpublished} 
lattices, all of those systems exhibiting local constraints and pinch-point singularities. From these considerations, 
it follows that this anisotropic spectral weight could be an intrinsic feature of Coulomb phases, a hypothesis that 
will have to be confirmed in further theoretical studies.
%

%%%%%%%%%%%%%%%%%%%%%%%%%%%%%%%%%%%%%%%%%%%%%%

\begin{figure}[t]
\includegraphics[width=8cm]{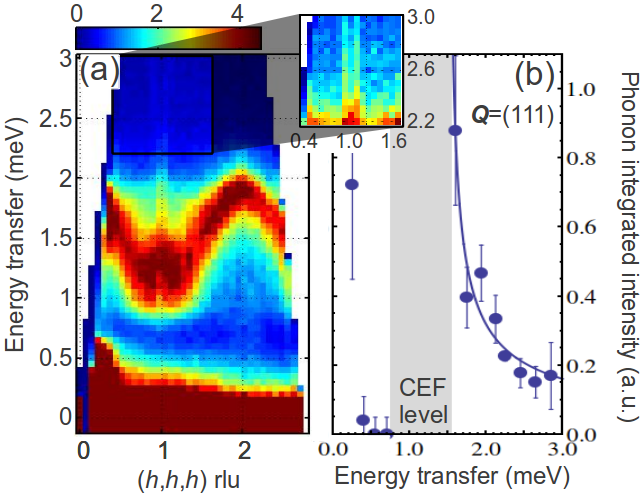}
\caption{(Color online) : (a) IN5 data showing the inelastic scattering as a function of energy 
transfer $\omega$ and $Q$ along $\langle h,h,h\rangle$. The data have been taken at 1.5~K 
but similar results are observable at 10 K. The crossing of the acoustic phonon mode and 
the dispersing CEF occurs close to $Q=(1,1,1)$. 
The inset has been plotted using a different intensity scale (from 0 to 1) to highlight the two 
branches of the acoustic phonon dispersion. (b) Simultaneously, the intensity of the phonon 
as a function of energy is strongly suppressed below the CEF line, and recovers an usual 
behavior above.}
\label{fig4}
\end{figure}

At slightly larger energies, $\omega= \Delta \sim$ 1.5 meV, the inelastic response is dominated by the 
first CEF excitations \cite{footonote2}. Since $\Delta$ is small, especially compared to classical spin ices 
(where $\Delta \sim$ 20 to 30 meV), the first CEF level is expected to play a significant role in the low 
energy properties of the system \cite{gingras,mirb07,kao}.
The line shape of this CEF excitation is much more complicated than a single dispersion less mode and 
very likely contains two different modes (not shown). It is strongly modulated at 10~K and down to the 
base temperature of 50~mK, because of the interactions between \tb\, magnetic moments \cite{kao}.
In a very narrow range of scattering vectors $\mathbf{Q}$ close to crystalline zone centers, such as 
$(1,1,1)$ and $(2,2,0)$, an unexpected upturn of the dispersion is observed (\ref{fig4}(a)). This upturn 
arises within 
the region of reciprocal space where there is a crossing between the crystal field level and the acoustic 
phonon branch stemming from the zone centers. Here, the phonon and the CEF seem to repel each 
other. To further illustrate this point, different cuts along $\langle h,h,h \rangle$ have been taken at 
different energy transfers from 0 to 3 meV. Figure \ref{fig4}(b) shows the corresponding 
$\mathbf{Q}$-integrated intensity of the phonon plotted as a function of $\omega$. In classical cases, 
it simply scales as $1/\omega$; in the present case however, a suppression of the phonon intensity below
 the CEF is observed. These features are the sign of a strong magneto-elastic coupling, although, here, 
the CEF level and the acoustic phonon does not seem to follow conventional 
hybridization processes \cite{gehring,pralo3,ceal2}. 

%%%%%%%%%%%%%%%%%%%%%%%%%%%%%%%%%%%%%%%

The issue remains to relate low energy propagating excitations and the strong magneto-elastic coupling. 
The existence of the former 
is indeed an intriguing question: because of the intrinsic properties of non-Kramers magnetic doublets, there are 
no matrix elements between the time conjugate states of the doublet $|\pm\rangle$ \cite{Mueller1967}, leading 
to a neutron cross section $~|\langle + | \hat{\mathbf{J}} |- \rangle|^2=0$. Non-zero matrix elements might 
in principle be restored by including the first excited CEF level \cite{molavian,Curnoe2008}. However, as long as 
the exchange terms are one order of magnitude weaker than $\Delta$, the perturbed wave function should not 
depart too much from $|\pm\rangle$, thus resulting in a vanishingly small inelastic spectral weight 
\cite{bonv11,petit-tbsn}.

To recover a significant cross section, it is therefore essential to go beyond a dipolar Hamiltonian, and to consider 
for example a coupling between quadrupolar moments \cite{Curnoe2008,petit}. In this respect, the magneto-elastic 
coupling responsible for the phonon and the CEF anomalies (see Fig. \ref{fig4}) could be the driving force leading 
to effective interactions between quadrupoles \cite{gehring}. There are additional clues in favor of a strong dynamical 
spin-lattice coupling: structural fluctuations below 15~K observed by high resolution X diffraction \cite{ruff}, giant 
magneto-striction \cite{ruff2} and the instability of the spin liquid state versus pressure and stress \cite{mirebeau-press}, 
all of which have been reported recently, but no static distortion has been observed so far \cite{goto}.

A model based on the most simple on-site quadrupolar term has been proposed, phenomenologically 
connected with a possible static tetragonal distortion precursor to a $T\simeq 0$ Jahn-Teller transition 
\cite{Curnoe2008,chapuis,mams,bonv09,bonvicfcm,bonv11,petit,petit-tbsn}. Despite being rather successful 
in explaining a number of experimental results \cite{pierre,petit,petit-tbsn}, it does not, in its present form, 
capture the whole nature of the ground state; for instance, it leads to a CEF singlet state on each site, which 
is not compatible with the existence of elastic correlations (see figure \ref {fig1}). Finding a more 
appropriate set of quadrupolar terms might be achieved on the basis of recent pseudo spins half effective 
models \cite{onoda,sungbin}. However, since the low energy branch is not the predicted emergent photon 
\cite{shannon}, the suitability of this approach to model \tbti\ remains unclear. Models based on several 
gauge fields \cite{Khemani2012} to account for the role of transverse spin components could be better suited, 
but the coupling between the $4f$ quadrupolar moments should definitely be considered. 

In summary, our neutron results demonstrate the existence of a low energy propagating excitation 
emanating from the spin liquid ground state of \tbti. Its peculiar spectral weight distribution could 
be the signature of propagating defects breaking the 
divergence-free flux characteristic of Coulomb phase. We also report anomalies of the phonon 
modes, as well as of the first CEF level, suggesting a strong dynamical coupling to the lattice. 
These experimental findings emphasize the importance of quadrupolar interactions in the physics 
of non-Kramers ions based quantum spin ices.

Authors acknowledge fruitful discussions with M. Gingras, B. Canals, E. Lhotel and A. Goukassov. 
We also acknowledge F. Damay for a careful reading of the manuscript.

%%%%%%%%%%%%%%%%%%%%%%%%%%%%%%%%%%%%%%%%%%%%%%%%%%%%%%%%%%%


\begin{thebibliography}{99}
\bibitem{footonote2} Because of the nominal trigonal symmetry at the rare earth site, the 
ground and first excited CEF state of the \tb\, ion are two doublets \cite{gingras,mirb07,kao}.
\end{thebibliography}
\end{document}